# Micropatterned Electrostatic Traps for Indirect Excitons in Coupled GaAs Quantum Wells

A. Gärtner[1], L. Prechtel[1], D. Schuh[2,3], A.W. Holleitner[1,*], and J.P. Kotthaus[1]

[1] Department für Physik and Center for NanoScience (CeNS), Ludwig-Maximilians-Universität München, Geschwister-Scholl-Platz 1, 80539 München, Germany.

[2] Institut für Angewandte Physik und Experimentelle Physik, Universität Regensburg, Universitätsstraße 31, D-93040 Regensburg, Germany.

[3] Walter Schottky Institut, Technische Universität München, Am Coulombwall 3, 85748 Garching, Germany.

We demonstrate an electrostatic trap for indirect excitons in a field-effect structure based on coupled GaAs quantum wells. Within the plane of a double quantum well indirect excitons are trapped at the perimeter of a $SiO_2$ area sandwiched between the surface of the GaAs heterostructure and a semitransparent metallic top gate. The trapping mechanism is well explained by a combination of the quantum confined Stark effect and local field enhancement. We find the one-dimensional trapping potentials in the quantum well plane to be nearly harmonic with high spring constants exceeding 10 keV/cm².

*e-mail: holleitner@lmu.de

**Introduction:**

The seminal work of Keldysh and Kozlov, predicting the possibility of Bose-Einstein condensation of excitons, has triggered many experiments aiming to observe the bosonic nature of excitons in solid state systems.[1] For detecting the Bose-Einstein condensation of excitons, it is a prerequisite to define controllable confinement potentials for excitons. So far trapping of excitons has been demonstrated in strained systems,[2],[3],[4] magnetic traps,[5] "natural traps" defined by interface roughness fluctuations,[6] and electrostatic traps.[7],[8],[9],[10] Electrostatic traps generally make use of the quantum confined Stark effect, which allows tuning the energy of excitons in quantum well structures by means of an electric field.[11],[12] Electrostatic traps combine advantages of other methods such as fast in-situ tunability, the creation of steep harmonic trapping potentials, and a large degree of freedom in varying the shape of the trap.[10],[13] In addition, electrostatic traps can be extended towards optoelectronic solid-state devices because of their potential scalability and compatibility with existing semiconductor technology.[14]

Here we report on a novel electrostatic trap, which gives rise to a very steep harmonic trapping potential for indirect excitons in one dimension. The trapping mechanism relies on a local electrostatic field enhancement in combination with the quantum confined Stark effect. The indirect excitons are trapped in GaAs quantum wells just below the perimeter of $SiO_2$-layers, which are sandwiched between the surface of the GaAs heterostructure and a semitransparent metallic top gate. We explain the exciton trapping via the electrostatic influence of surface states at the $GaAs/SiO_2$ interface. We find nearly harmonic trapping potentials with spring constants of ~10 keV/cm². The value exceeds previous results on coupled quantum wells by a factor of 300.[4],[13] Such electrostatic traps for indirect excitons may ultimately be exploited for hosting an excitonic Bose-Einstein condensate.[15],[16],[17]

**Experiment:**

Starting point is an epitaxially grown AlGaAs/GaAs heterostructure, which contains two GaAs quantum wells encompassed by $Al_{0.3}Ga_{0.7}As$ barriers (see Fig. 1(a)). Each quantum well has a thickness of 8 nm, and the quantum wells are separated by an $Al_{0.3}Ga_{0.7}As$-tunnel barrier with a thickness of 4 nm. The center of the double quantum wells is located 60 nm below the surface of the heterostructure. An n-doped GaAs-layer at a depth of $d = 370$ nm serves as a back gate, while a semi-transparent titanium-layer is used as the top gate of the field effect device. In coupled quantum wells embedded in a field-effect structure, electrons and holes of photogenerated excitons may rearrange in a way that they are spatially separated by the tunnel barrier between the GaAs quantum wells.[18] Such indirect excitons have a long lifetime, which is electrically tunable and which reaches values of up to 30 μs.[20],[21] In contrast, the optical lifetime of direct excitons in quantum wells is shorter than 1 ns (for T = 5 K). As depicted in Fig. 1(a) and (b), the investigated samples feature an additional $SiO_2$-layer, which is sandwiched between the GaAs surface and the metal top gate. The thickness of the $SiO_2$-layer is about 50 nm, and the titanium top gate has a thickness of about 10 nm. The top gate can be distinguished in two regions: the bias gate, which is in direct contact with the GaAs heterostructure, and the control gate, which is located on top of the $SiO_2$-layer. As there is no electrical connection between the bias gate and the control gate, the two regions can be tuned independently to different voltages $V_B$ and $V_C$.[23] All experiments on exciton trapping and storage are carried out in a continuous-flow helium cryostat at a temperature of 3.8 K in combination with a time-resolved micro-photoluminescence setup. The excitons are locally excited within the coupled quantum wells by focusing a pulsed laser onto the sample (see Fig. 1(a) and (b)). The laser wavelength of 680 nm is chosen such that the corresponding energy $E_{Photon}$ = 1.82 eV is above the effective band gap of the GaAs quantum wells and below the band gap of the $Al_{0.3}Ga_{0.7}As$ barriers.[24] The laser is operated at a pulse length of 50 ns, and the repetition period is set to be 10 μs. The laser beam is focused to a spot with a diameter of 10 μm, and a typical corresponding power density is 5 kW/cm². The photoluminescence signal of the recombining excitons is collected by a optical microscope and studied as a function of the delay time with respect to the initial laser pulse. The optical signal is analyzed by a triple-grating imaging spectrometer with an energy resolution of ~0.2 meV and a lateral spatial resolution of ~2 μm. A fast-gated, intensified charge coupled device (ICCD) camera with an exposure time of 5 ns detects the excitonic photoluminescence as a function of energy and position. In order to obtain a sufficient signal to noise ratio, the experiments are performed by integrating over 4 x $10^6$ single events. Fig. 1(c) shows a typical photoluminescence pattern of excitons recombining along the edge of a $SiO_2$-layer both with convex and concave curvatures at a delay time of 3 μs. The excitonic photoluminescence pattern occurs just at the position dividing the bias and the control gate, implying that the recombining excitons are trapped at the perimeter of the $SiO_2$-layer. A delay time of up to 10 μs between the laser excitation and the detection of the excitonic photoluminescence allows long-living indirect excitons to diffuse a distance of several hundreds of microns along the line-shaped trap with respect to the excitation laser spot. Since we find the long-living excitons to propagate a distance of about 300 µm in less than the time resolution of our experiment of about 10 ns, we estimate the spreading to exceed $(2-3) \times 10^4$ m/s. This spreading found in the perimeter trap exceeds by far diffusion and drift dynamics observed in devices without the $SiO_2$-induced trap.[18],[19] Experiments with higher time resolution aim at measuring and understanding the spreading dynamics in more detail (data not shown).

In Fig. 2(a) and (b), we demonstrate that the voltage $V_C$ controls the excitonic photoluminescence pattern in the region between the bias and the control gate. The delay time between laser excitation and photoluminescence detection is chosen to be 800 ns, while the exposure time of the recording ICCD camera is 50 ns. The distance between the spot of laser excitation and the $SiO_2$-layer is set to be about 50 µm.[25] In Fig. 2(a) [(b)], the control voltage is set to be $V_C = 0$ V [$V_C = 0.25$ V], while the bias voltage is fixed at $V_B = 0$ V. Remarkably, only for $V_C = 0$ V, the ring-shaped photoluminescence emission is observed, which is located just outside the perimeter of the $SiO_2$-layer (dashed line).[26] In Fig. 2(c) a typical time-integrated photoluminescence signal, collected from the area of the bias gate, is depicted, while the data in Fig. 2(d) are collected from the area of the control gate. Indirect excitons are only observed beneath the bias gate [black arrow in Fig. 2(c)]. The spectral position of indirect excitons can be shifted by the quantum confined Stark effect, as recently reported.[18] The photoluminescence maximum at about 785 nm in both graphs corresponds to direct excitons. Since the lifetime of direct excitons is shorter than 1 ns at T = 5K, the latter can only be detected at short delay times. In addition, the spectral position of the direct excitons excited below the $SiO_2$ in Fig. 2(d) is independent of both the control and the bias gate (data not shown). This observation can be explained by the presence of a large density of states formed at the surface of GaAs crystals.[27] The surface states, originating from dangling bonds, are energetically located in the middle of the GaAs band gap. It is well known that the surface states form a Schottky barrier at the metal – GaAs interface.[28] Consequently, we observe flat-band conditions in our samples at bias voltages of roughly $V_B$ = +0.8 V, corresponding to the Schottky barrier. However, the situation changes when the GaAs surface is covered with a $SiO_2$ layer. As $SiO_2$ is electrically insulating, photo-excited charge carriers cannot be drained and, consequently, they are accumulated in unoccupied surface states at the GaAs/$SiO_2$ interface. The charge accumulation process continues until the resulting sheet charge density at the surface screens the external electric fields and flat-band conditions arise in the plane of the coupled quantum wells. Due to flat-band conditions beneath $SiO_2$-covered regions, only direct excitons can exist there, and their spectral position is independent of the applied control gate voltage $V_C$ (Fig. 2(d)).[26]

In Fig. 3(a), the laterally integrated total intensity of the photoluminescence emitted from the circular trap in Fig. 2(a) and (b) is plotted as a function of the applied control voltage $V_C$, while the bias gate voltage is fixed at $V_B$ = +0.3 V. A time delay of 800 ns ensures that only indirect excitons are recorded. Two regimes are distinguishable: at a low control voltage, $V_C$ < 0.48 V, the intensity of the detected photoluminescence is high (gray shaded region) and independent of the applied control voltage. For $V_C$ > 0.48 V the intensity drops abruptly to the background noise level within $\Delta V_C$ ~ 30 mV. Within the experimental resolution we do not observe a hysteresis when crossing $V_C$ ~ 0.48 V.[29] Most importantly, the trapping behavior depends also on $V_B$. Fig. 3(b) shows a phase diagram of the trapping behavior as a function of both $V_C$ and $V_B$. The phase boundary (diagonal solid line), which separates the trapping state from the non-trapping state, can be described by following empirical formula: $V_C = 0.784 \cdot V_B + 0.245\ V$ (3). Note that the actual state of the trap is solely determined by the control and the bias voltages. At the same time, the characteristic of the phase boundary is a clear signature of an electrostatic origin of the trapping mechanism.

**Model:**

We devise a phenomenological electrostatic model to describe the origin of the excitonic trapping potential. The model is based on two major assumptions. First, as described above, there exists a large density of charged surface states[27],[28] at the vertical interface between the $SiO_2$-layer and the GaAs-surface such that the electric field perpendicular to the double quantum well becomes negligibly small.[26] Second, the area around the perimeter of the $SiO_2$-layer can be divided into three regions I, II, and III, each one being at constant electrostatic potentials $V_I$, $V_{II}$, and $V_{III}$ at the GaAs surface, respectively. Fig. 4(a) shows a top view image as well as a cross sectional view of such a fragmentation. Region I denotes the area beneath the $SiO_2$-layer with a constant potential independent of the control gate voltage $V_C$. Region III corresponds to the area below the bias gate with a constant electric field controlled by $V_B$. Region III is assumed to start at a minimum distance of 5 μm from the $SiO_2$-layer. Region II is defined as the transitional area between regions I and III. On one hand, the actual "intermediate" electric potential in region II is dominated by the bias voltage $V_B$ because region II is also covered by the bias gate. On the other hand, the potential $V_{II}$ is strongly influenced by the close vicinity of the surface charge density $\rho_{surface}$ in region I causing an effective field enhancement. Using finite element techniques,[30] we calculate the electrostatic potential $\varphi(r)$ within the heterostructure in accordance to the phenomenological model (Fig. 4(a)). The energy of indirect excitons $U_{exc}$ at the depth of the coupled GaAs quantum wells is governed by the vertical electric field via the quantum confined Stark effect,

$$\Delta U_{exc}(\boldsymbol{r}) = -e \cdot d_{eff} \cdot \left\| \frac{d\varphi(\boldsymbol{r})}{dz} \right\|_{z=z_0}, \qquad (1)$$

where $\boldsymbol{r}$ is the spatial position vector, $z_0$ is the position of the center of the coupled quantum wells along the growth direction $z$, $d_{eff} \sim 10$ nm is the effective separation of electrons and holes forming the indirect excitons, and $e$ is the electron charge. Fig. 4(b) shows the energy landscape of excitons $\Delta U_{exc}(r_\parallel)$ deduced via Eq. (1) from the calculated electrostatic potential $\varphi(r)$ shown by contour lines in Fig. 4(a). The calculations are done for different values of the intermediate electric potential $V_{II}$, whereas the electrostatic potentials of regions I and III are set to be $V_I = -0.78$ V and $V_{III} = 0$V. For $V_I < V_{II}$, the numerical calculations show a minimum in the excitonic energy. The minimum is laterally located at the position $x \sim -60$ nm with respect to the boundary between regions II and III (see Fig. 4(c)). For $V_I > V_{II}$, the indirect excitons are expected to reside in the minimum at the position x $\sim$ +50 nm (see Fig. 4(b)).[31] Experimentally, this lateral displacement of the trap minimum of about ~100 nm cannot be resolved with the present micro photoluminescence setup. The energy minimum traps the indirect excitons and is caused by a combination of an effective field enhancement in vertical direction and the quantum confined Stark effect. The lateral component of the electric field $E_{lat}$ in the coupled quantum wells is shown in the inset of Fig. 4(b) for an increasing intermediate electrostatic potential $V_{II}$. Exciton ionization takes place as soon as the lateral electric field $|E_{lat}|$ exceeds a value of approx 1 V/μm.[13],[32],[33] For $V_{II} > -0.6$ V, the ionization level at the lateral position of the trap is exceeded (dashed curve). Consequently, excitons are ionized into spatially separated electrons and holes, and excitonic luminescence is quenched as seen in Fig. 2(b). Experimentally, the vertical field enhancement can be seen by the fact that the exciton lifetimes differ inside and outside the trap. We find that indirect excitons in the trap have lifetimes exceeding 10 μs, which is typically a factor 100 longer than the lifetimes of indirect excitons in region III far away from the trap. In addition, samples with $SiO_2$-layers, which are deposited on top of a uniform bias gate, do not show any trapping of indirect excitons (data not shown). In that case, the metal gate fixes the surface potential, and no field enhancement effects occur at the perimeter of the $SiO_2$-layer. Note that the transversal width of the trap was measured to be roughly 10 μm (see Fig. 2(a)). This is in contrast to the value of 100 nm predicted by the model. We attribute this finding to the following facts. First, the excitons may spill out of the center of the trap into region III. This is reasonable as the trap is not limited by a potential barrier towards region III, and high laser excitation powers were employed.[34] Moreover, the calculations were not performed self-consistently neglecting screening effects by mobile charge carriers. Mobile charge carriers provide further screening of the electric field constituting the trap, giving rise to a further flattening of the trapping potential. Nevertheless, it is a great success of this model to give a semi-quantitative explanation of the trap behavior.

**Discussion:**

In order to characterize the trapping potential and to further test the electrostatic model presented above, we mapped the central energy of the indirect excitons versus the lateral coordinate $x$. The inset of Fig. 5(b) depicts the experimental setup. A square $SiO_2$-layer forms an exciton trap along its perimeter. The trap is populated with excitons by a focused laser beam. The laser spot is located in direct vicinity to the $SiO_2$-layer in order to maximize the exciton population of the trap. The bias voltage is set to $V_B = 0.3$ V and the control gate is floating. Around two positions, (a) and (b), the photoluminescence at a time delay of 50 ns is simultaneously resolved in space (dashed lines) and energy via the imaging spectrometer. The spatial distribution of the excitonic energy $U_{exc}$ at both positions is shown in Fig. 5(a) and (b), with $x$ being the lateral position. The cross section of the sample at each position is sketched on top of both figures. The energy minima are located outside of the $SiO_2$-layer's perimeter in agreement with the electrostatic model. The solid curves represent harmonic functions, $U_{exc} = ½ \, kx^2$, which are used to approximate the data points depicted in Fig. 5(a) and (b), with $k$ denoting the spring constant. For the trapping potentials shown in Fig. 5(a) and (b), spring constants of 11 keV/cm² and 1.1 keV/cm² are yielded. The spring constants exceed values obtained by other methods by up to 300 times.[16] The spring constant and the position of the trap minimum do not depend on the laser power within the experimental error (data not shown).[35] Steep confining potentials are necessary to create dense droplets of excitons leading, ultimately, to a Bose-Einstein-condensate (BEC) of excitons. In terms of the quantization energy of a particle in a harmonic potential $\eta\omega = \eta\sqrt{k/m^*}$, with the exciton mass $m^* = 0.25\, m_0$ in GaAs and the free electron mass $m_0$, the spring constants correspond to 5.5 µeV and 1.7 µeV, respectively. Both values are too small to resolve quantization of the excitonic energies with the experimental methods applied here. However, assuming a classically defining, quasi zero-dimensional trap with a spring constant $k$ of 11 keV/cm², a critical transition temperature $T_C$ as high as 2 K is expected, even at a comparable low exciton density of $10^9$/cm².[1],[15],[16],[17] Generally, we attribute the observed difference of spring constants $k$ in Fig. 5(a) and (b) to possible asymmetries during the sample fabrication, leading to local field inhomogeneities. For instance, during the deposition of the $SiO_2$-layer, thickness fluctuations at the perimeter of the structures are probable. At the same time, dipole-dipole interactions can change the energy of the recombining excitons as a function of the excitonic density in the trap.[36],[37],[38] However, since the spring constants do not depend on the laser intensity, dipole-dipole interactions are considered to play only a minor role for the shape of the confinement potential.

Another possibility to form potential landscapes for excitons is the application of mechanical strain to semiconductor samples, as described by the Pikus and Bir deformation Hamiltonian.[4],[16] Our samples comprise three different materials: (Al)GaAs, $SiO_2$, and thin metal films. Due to the different thermal expansion coefficients, the semiconductor can be assumed to be strained mechanically, when the samples are cooled down to cryogenic temperatures. We numerically modeled the effect of mechanical strain on excitons in the whole heterostructure using the software package nextnano$^3$.[39] To this end, the effect of stress is estimated in a way that the lattice constant of the semiconductor surface is stretched or compressed by up to 20 % beneath a stressor-layer. All calculations for the band energy of the GaAs quantum wells show the same qualitative result (as depicted in Fig. 6(a)): a strongly anisotropic shape of the trapping behavior with respect to the crystal axis. Exciton confinement should mainly occur in the [110] direction, but not in [-110] direction. A typical photoluminescence pattern for a circular $SiO_2$-layer at low temperatures is shown in Fig. 6(b). Obviously, no anisotropy is detected in the data. We performed the experiment at different laser powers to make sure that the predicted strain effect on excitons is not overwhelmed by excitons spilling out from the trap (data not shown). Since the experimental finding is in contrast to the predictions of a strain model, we conclude that strain effects are negligible in our experiment. In addition, we would like to note that the concave and convex curvatures of the $SiO_2$-layer in Fig. 1(c) do not influence the trapping behavior of the device.

**Summary:**

In summary, we report on the successful realization of an electrostatically tunable trap, which gives rise to a harmonic trapping potential for indirect excitons in one dimension of the quantum well plane. The harmonic trapping potentials exhibit spring constants of ~10 keV/cm². The steep confinement potential for excitons in conjunction with the in-situ tunability of the trap promotes the technique to a promising candidate for future experiments on Bose-Einstein condensation of excitons in a fully confining quasi zero-dimensional potential landscape. Future work aims towards realizing such traps.

We thank A.O. Govorov for very fruitful discussions on the electrostatic origin of the exciton trap. We gratefully acknowledge financial support by the grant KO-416/17 of the Deutsche Forschungsgemeinschaft (DFG), the Center for NanoScience (CeNS) in

Munich, and the German excellence initiative via the cluster "Nanosystems Initiative Munich (NIM)".13

**Figures:**

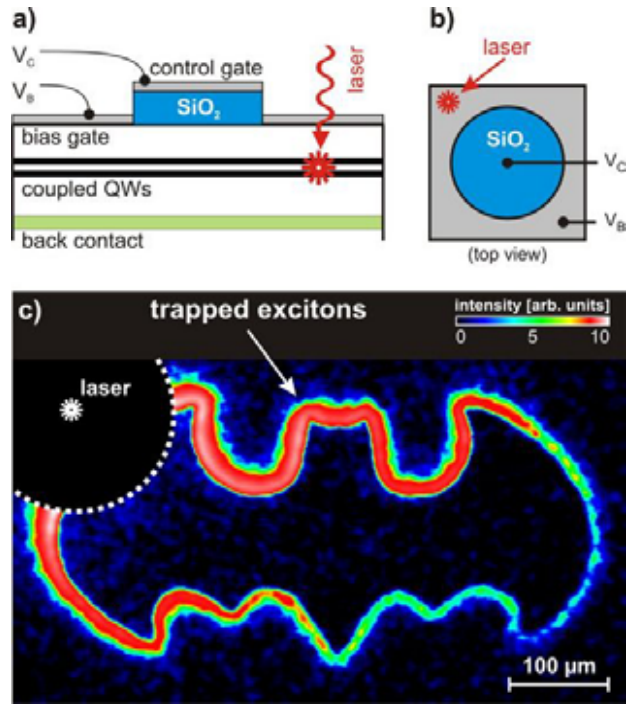

**(Color online) Figure 1: (a) Schematic side view and (b) top view of the field effect structure with an additional $SiO_2$-layer on top of the GaAs/AlGaAs heterostructure. Long-living excitons are laser-generated in the coupled quantum wells (QWs). By appropriately tuning the bias $V_B$ and the control voltage $V_C$ with respect to the back gate, excitons can be captured and stored along the perimeter of $SiO_2$-layers with varying curvature as shown in (c). The intensity of the photoluminescence is color-coded, while the bias voltage is set to be $V_B = 0$ V and the control gate is floating.**

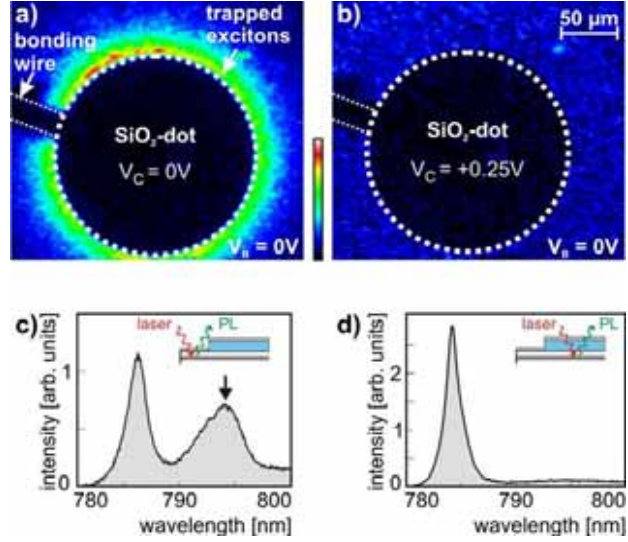

**(Color online) Figure 2: Excitonic photoluminescence emitted from an electrostatic, circular trap at $T = 3.7$ K. After the excitons have been generated by a pulsed laser approximately 50 μm away from the circular trap, they diffuse into the trap and recombine. The trap can be electrically switched from "on" (a) to "off" (b) by voltage $V_C$, while voltage $V_B$ is kept constant at 0 V. The dashed circle depicts the perimeter of the circular SiO$_2$-layer with a diameter of 160 μm. By using a band-pass filter we ensure that only photoluminescence of indirect excitons is recorded. (c) Time-integrated photoluminescence on top of the bias gate at a voltage of $V_B = 0$ V without a filter, showing both direct and indirect (arrow) excitons. (d) Equivalent measurement on top of the control gate exhibits only direct excitons. See text for details.**

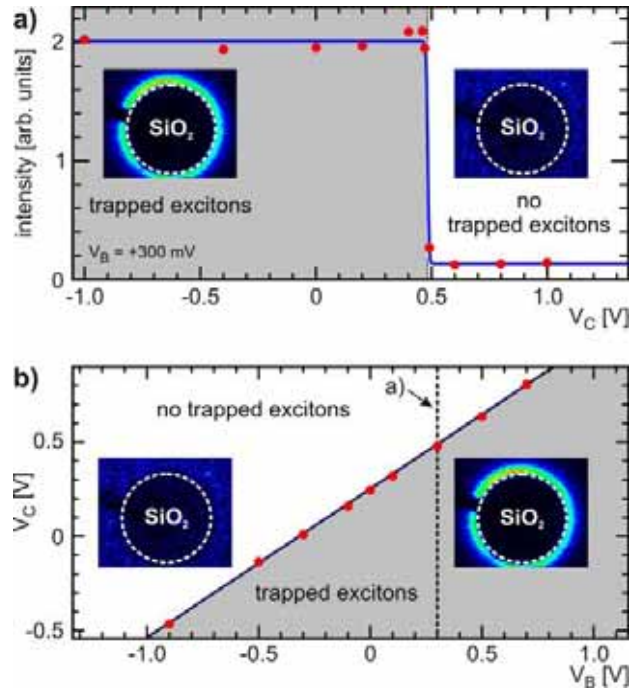

**(Color online) Figure 3: Switching characteristics of the circular trap. (a) At a constant bias voltage of $V_B = +0.3$ V, the circular excitonic photoluminescence pattern is detected for $V_C < 0.48$ V. The solid curve is a guide to the eyes. (b) Phase diagram of the trapping behavior in the $V_C$-$V_B$ space. The dashed line refers to the data shown in (a).**

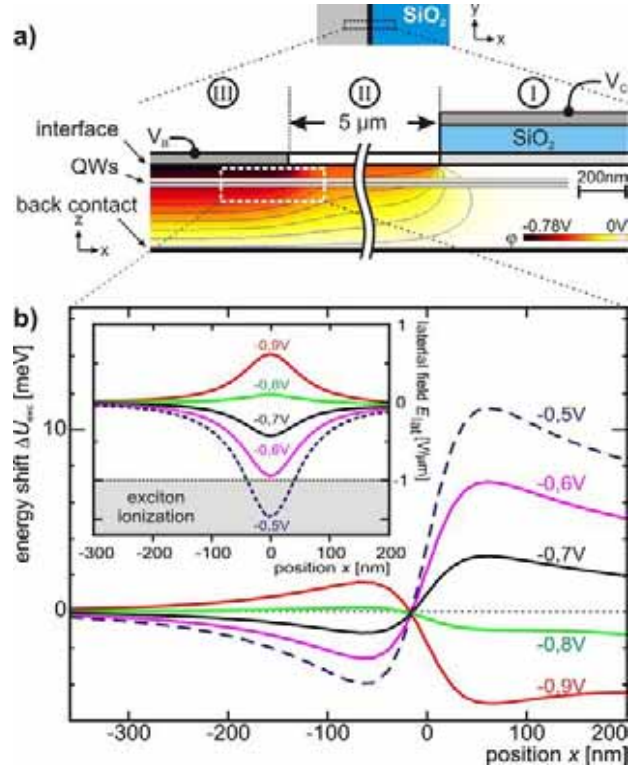

(Color online) Figure 4: (a) Electrostatic potential $\varphi(r)$ (color-coded) at the perimeter of the SiO$_2$-layer (see inset on the top) for $V_I = 0$ V, $V_{II} = -0.4$ V, and $V_{III} = -0.78$ V. The thin curves indicate equipotential lines of the electrostatic potential. (b) Calculated excitonic energy $\Delta U_{exc}$ due to the quantum confined Stark effect in the region indicated by the dashed rectangle in (a). The position $x = 0$ corresponds to the boundary between region II and III, where a minimum of excitonic energy is found. Inset: increasing the voltage $V_{II}$ above approximately -0.6 V induces lateral electric field components $E_{lat}$, which eventually give rise to exciton ionization.

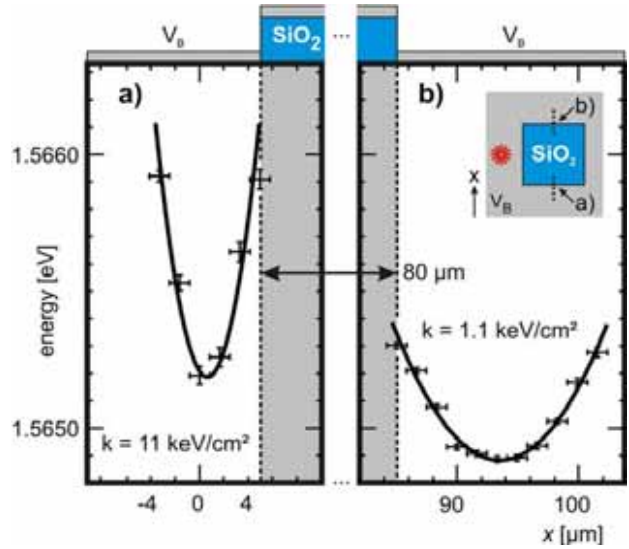

**(Color online) Figure 5: Excitonic recombination energy $U_{exc}$ as a function of the position $x$. The excitons are trapped a few microns away from the perimeter of the SiO$_2$-layer at $T = 3.7$ K (see schematic top view in the inset of (b)). The spring constants $k$ of the trapping potentials shown in (a) and (b) are obtained by numerically fitting harmonic functions (solid curves) to the measured data.**

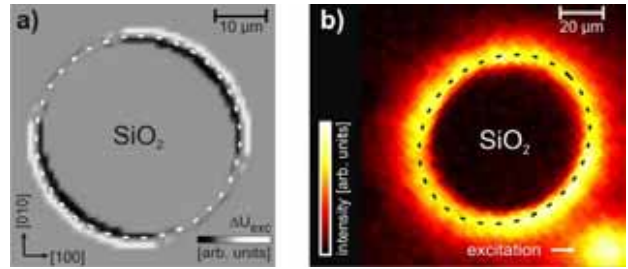

**(Color online) Figure 6: (a) Predicted excitonic energies in coupled quantum wells located 30 nm below a surface stressor. Here, the exciton trapping is expected to be anisotropic, because boundaries of the stressor facing the [110]-direction give rise to lower exciton energies than the ones in the [-110] direction. (b) Measured photoluminescence emission of a circular trap at $T$ = 3.7 K. Strain effects can be excluded as a dominant effect for trapping, since the photoluminescence is isotropic along the boundaries of the SiO$_2$-layer.**